\documentclass[useAMS,usenatbib]{mn2e}
\usepackage{url,times,graphicx,amsmath,amsfonts,amssymb,aas_macros,color,epsfig,varioref,subfigure,comment,natbib,appendix}
\usepackage{longtable}

\newcommand{\Tab}[1]{Table~\ref{#1}}
\newcommand{\Sec}[1]{Section~\ref{#1}}
\newcommand{\App}[1]{Appendix~\ref{#1}}

\newcommand{\Fig}[1]{Fig.~\ref{#1}}
\newcommand{\hMpc}{{\ifmmode{h^{-1}{\rm Mpc}}\else{$h^{-1}$Mpc}\fi}}
\newcommand{\hGpc}{{\ifmmode{h^{-1}{\rm Mpc}}\else{$h^{-1}$Gpc}\fi}}
\newcommand{\hkpc}{{\ifmmode{h^{-1}{\rm kpc}}\else{$h^{-1}$kpc}\fi}}

\newcommand{\hMsun}{{\ifmmode{h^{-1}{\rm {M_{\odot}}}}\else{$h^{-1}{\rm{M_{\odot}}}$}\fi}}
\newcommand{\Msun}{{\ifmmode{{\rm {M_{\odot}}}}\else{${\rm{M_{\odot}}}$}\fi}}
\def\hMpc{$h^{-1}\,{\rm Mpc}$}
\def\hkpc{$h^{-1}\,{\rm kpc}$}
\def\kpc{\rm kpc}
\def\Mpc{\rm Mpc}
\def\kms{{ \rm km} $s^{-1}$}
\def\LCDM{\ensuremath{\Lambda}CDM}
\def\vtan{$v_{\rm tan}$}
\def\vtanI{$v_{\rm tan}^{(I)}$}
\def\vtanII{$v_{\rm tan}^{(II)}$}
\def\vrad{$v_{\rm rad}$}
\def\mlg{$M_{\rm LG}$}
\def\std{Rand}
\def\modo{Mod1}
\def\modtw{Mod2}
\def\modth{Mod3}
\def\modfo{Mod4}
\def\modfi{Mod5}
\def\mods{Mod6}
\def\mvir{$M_{\rm vir}$}
\def\mth{$M_{\rm 200}$}
\def\mmw{$M_{\rm MW}$}
\def\mmto{$M_{\rm M31}$}

\title
[Constraining the mass of the Local Group]
{Constraining the mass of the Local Group}
\author[Carlesi Edoardo]
{
Edoardo Carlesi, $^{1}$
\thanks{E-mail: carlesi@phys.huji.ac.il}
Yehuda Hoffman,$^{1}$
Jenny G. Sorce,$^{2}$
Stefan Gottl\"ober$^{2}$\\
$^{1}$Racah Institute of Physics, Givat Ram, 91040 Jerusalem, Israel\\
$^{2}$Leibniz-Institut f\"ur Astrophysik Potsdam (AIP), An der Sternwarte 16, D-14482 Potsdam, Germany}

\begin{document}

\date{Submitted XXXX May XXXX}

\pagerange{\pageref{firstpage}--\pageref{lastpage}} \pubyear{2016}

\maketitle

\label{firstpage}

%%%%%%%%%%%%%%%%%%%%%%%%%%%%%%%%%%%%%%%%%%%%%%%%%%%

\begin{abstract}
The mass of the Local Group (LG) is a crucial parameter  for galaxy formation theories.
However, its observational determination is challenging - its mass budget is dominated by dark matter which cannot be directly observed. 
To meet this end the posterior distributions of the LG and its massive constituents have been constructed by means of constrained 
and random cosmological simulations. 
Two priors are assumed - the \LCDM\ model that is used to set up the simulations and a LG model, 
which encodes the observational knowledge of the LG and is used to select LG-like objects from the simulations. 
The constrained simulations are designed to reproduce the local cosmography as it is imprinted onto the Cosmicflows-2 database of velocities. 
Several prescriptions are used to define the LG model, focusing in particular on  different recent estimates of the tangential velocity of M31. 
It is found that 
(a) different \vtan choices affect the peak mass values up to a factor of 2, and change mass ratios of \mmto\ to \mmw\  by up to 20\%; 
(b) constrained simulations yield  more sharply peaked posterior distributions compared with the random ones; 
(c) LG mass estimates are found to be smaller than those found using the timing argument; 
(d) preferred MW masses lie in the range of $(0.6 - 0.8)\times10^{12}$\Msun\ whereas 
(e) \mmto\ is found to vary between $(1.0 - 2.0)\times10^{12}$\Msun, with a strong dependence on the \vtan values used. 
\end{abstract}

%%%%%%%%%%%%%%%%%%%%%%%%%%%%%%%%%%%%%%%%%%%%%%%%%%%%%%%%%%%%%%%%%%%%%%%%%%%%%%%%%%
%%%%%%%%%%%%%%%%%%%%%%%%%%%%%%%%%%%%%%%%%%%%%%%%%%%%%%%%%%%%%%%%%%%%%%%%%%%%%%%%%%

\section{Introduction}\label{sec:intro}

The impressive developments of cosmology of the past few years have been driven by a remarkable improvement in the quality and quantity of 
available data as well as a substantial effort in the theoretical and numerical modeling devoted at accommodating them within a single, 
coherent framework, that is the standard $\Lambda$ Cold Dark Matter model (\LCDM).
Thanks to the advancements in the numerical simulation techniques
it has become possible to simulate objects of the size of our Local Group (LG) of galaxies, 
allowing to test both standard \citep[e.g.][]{Boylan-Kolchin:2012, Zavala:2012, Garrison:2014, Tollerud:2014}
and alternative cosmological theories on small scales \citep[e.g.][]{Elahi:2015, Penzo:2016, Garaldi:2016}.
This is the subject of near field cosmology, whose purpose is to extract cosmologically-relevant information 
from the high-quality observations of the nearby universe.

However, estimating the abundance of dark matter (DM) in the immediate neighbourhood is still challenging.
In particular, the exact size of the two halos hosting the Milky Way (MW) and Andromeda (M31) galaxies is known 
only within a large degree of uncertainty.
Viral mass estimates for the DM haloes of both M31 and MW are still just loosely constrained by observations 
\citep{Kochanek:1996, Wilkinson:1999, Evans:2000, Marel:2008, Karachentsev:2009, Marel:2012, Gibbons:2014} and in general need
a substantial degree of modelling and theory to be inferred 
\citep[see e.g.][]{Peebles:2001, Li:2008, McMillian:2011, Fardal:2013, Boylan-Kolchin:2013, Diaz:2014, Cautun:2014, Wang:2015}.

A popular and straightforward way of addressing the problem is represented by the "timing argument" \citep[TA, ][]{Kahn:1959} which models the
LG halo pair as point-like particles that depart from each other after the Big Bang while decelerating under their own gravity, until 
they start approaching. 
Although simple and insightful, this simple analytical approach can be 
improved by further modeling including elements such as dark energy \citep{Partridge:2013} or the Large Magellanic Cloud \citep{Penarrubia:2016}.

Cosmological $N$-body simulations, on the other hand, provide an alternative tool to study the mass of 
the LG and its members \citep{Busha:2011, Gonzalez:2014}. 
However, they require a prior knowledge of the cosmological model and the objects under
investigation. This means that, in general, a suitable \emph{Local Group Model}, a set of prior that incorporates
our previous knowledge of the system, has to be introduced and specified.

In this \emph{Paper} we address the issue of determining the total mass of the local group \mlg\ and its main components
following the method described in \citet{Carlesi:2016b} to derive posterior distribution functions for the tangential velocity of M31 
using a simple LG model.
This kind of Bayesian approach is in turn similar to the what \citet{Busha:2011} used to constrain \mmw\
and to that of \citet{Gonzalez:2014} for the \mlg. 
In the present work, most of the emphasis is put on the consequences that
the different \vtan\ estimates of \citet{Sohn:2012} and \citet{Salomon:2016} (referred to as \vtanI\ and \vtanII\ hereafter) 
have on the mass of the Local Group of galaxies,  
and on the masses of the two largest DM haloes, M31 and MW. 
In fact, whereas the first authors directly measured an extremely small value
of \vtan, compatible with zero at the $1\sigma$ level, the latter found that new satellite data favour a much higher \vtan$\approx 160$\kms.
Each one of these values is expected to have substantially different implications for the mass of the LG and its members,
which are going to be discussed here.

Moreover, two different halo samples are used. 
The first one comes from the so called Local Group Factory simulation series
introduced by \citet{Carlesi:2016a}, and provides a large number of LG-like objects within a large scale environment 
quantitatively and qualitatively close to the real one. 
Here, the Cosmicflows 2 \citep[CF2, ][]{Tully:2013} dataset is used to generate the initial conditions, 
constraining the variance of the simulations and thus shaping the $z=0$ properties of the LG neighbourhood.
The second one is a control sample drawn from a standard (random) \LCDM\ simulation. 
The comparison between the results obtained within the two samples will allow to highlight the influence of the particular environment 
on \mlg, \mmto\ and \mmw, and disentangle it from the effect of the kinematic priors used to define LGs.

This work is structured as follows. \Sec{sec:methods} contains a description of the simulations and the methods used, 
with particular emphasis on the choice of the priors used to derive a posterior distribution functions.  
\Sec{sec:results} describes these result for \mlg, \mmto, \mmw\ and their ratio under different choices of the priors,
highlighting the influence of \vtan\ on the final estimates and also comparing the results to the values
provided by other authors. In \Sec{sec:conclusions} we summarize our main results and discuss their implications.

%%%%%%%%%%%%%%%%%%%%%%%%%%%%%%%%%%%%%%%%%%%%%%%%%%%%%%%%%%%%%%%%%%%%%%%%%%%%%%%%%%

\section{Methods}\label{sec:methods}
In this section we review the methods and the simulations used to derive the posterior likelihood functions,
as already discussed by \citet{Carlesi:2016b}.

\subsection{The Simulations}
LG-like halo pairs are drawn from two classes of \LCDM\ simulations running with Planck-I 
cosmological parameters $\Omega _m = 0.31$, $\Omega _{\Lambda} = 0.69$, $h=0.67$ and $\sigma_8 = 0.83$ \citep{Planck:2013}.
DM halo catalogues are produced using the \texttt{AHF} halo finder \citep{Knollmann:2009}; halo masses are defined by
$M_{200}$ with respect to $\rho_{crit}$.
Total LG mass is simply defined as \mlg = \mmw\ + \mmto.
The first LG sample is generated by selecting pairs out of 700 Constrained Simulations (CSs), 
obtained using the Local Group factory numerical pipeline \citep{Carlesi:2016a}.
It has been shown that these candidates live by construction in a large scale environment akin to the observational one, starting
from suitable Initial Conditions (ICs) \citep{Doumler:2013a, Doumler:2013b, Doumler:2013c, Sorce:2014, Sorce:2016} generated
using (grouped) CF2 peculiar velocity data \citep{Tully:2013} as constraints; biases have been minimized using the method of \citet{Sorce:2015}. 
The standard random realization of the \LCDM\ model, from which the second sample (labeled \std) was drawn, has been  
provided by the CurieHZ project \footnote{http://curiehz.ft.uam.es/}, and consists of a single DM simulation with $1024^3$ particles within a
200\hMpc\ box.
The mass resolution in both CS and \std\ simulation is $6.5\times10^{8}$\hMsun, which allows to resolve haloes of mass
$\sim 10^{12}$\hMsun with $\sim 200$ DM particles.
Thanks to the fact that both simulation types have the same DM particle mass, it is possible to factor out
numerical effects when comparing the results between them.

\subsection{The Local Group Model}

\begin{table}
\begin{center}
\caption{Kinematic priors on velocities (in km/s) and relative distances of the haloes (in \hMpc).
The first (last) three choices feature $\pm25\%$ ($\pm50\%$) intervals around the $r$ and \vtan\ values of \citet{Marel:2012}.
\modo\ features no \vtan\ restriction, for \modtw\ and \modfo\ \vtanI$\pm 1\sigma$ have been chosen while \modth\ and \mods\
use \vtanII$\pm1\sigma$ priors.}
\label{tab:priors}
\begin{tabular}{cccc}
\hline
\quad & $V_{rad}$ & $V_{tan}$ & $r$ \\
\hline
\modo\ & $[-135, -80]$ & $-$ 	  & $[0.35, 0.70]$ \\
\modtw & $[-135, -80]$ & $[0, 34]$ 	  & $[0.35, 0.70]$ \\
\modth & $[-135, -80]$ & $[103, 225]$ & $[0.35, 0.70]$ \\
\hline
\modfo & $[-150, -55]$ & $-$ 	  & $[0.25, 0.78]$ \\
\modfi & $[-150, -55]$ & $[0, 34]$ 	  & $[0.25, 0.78]$ \\
\mods & $[-150, -55]$ & $[103, 225]$ & $[0.25, 0.78]$ \\
\hline
\end{tabular}
\end{center}
\end{table}

Before discussing the properties of LG-like objects, it is obviously necessary to specify the
set of properties used to identify it.
This is what we call here \emph{Local Group model} - in different contexts (e.g. hydrodynamical simulations)
different models can be invoked.
Any choice of the Local Group Model reflects our prior observational knowledge on the system considered.
Following the observation that the LG system is dominated by a pair of spiral galaxies, the MW and M31,
simulated LGs are first of all defined as a pair of DM haloes.

M31 (MW) is assumed to be the most (least) massive of the two, as suggested by a host of different considerations such as
the measure of the tidal force acting on the MW \citep{Baiesi:2009},  
the position of the barycenter of the LG given by the local Hubble flow \citep{Karachentsev:2009} and
the balance of the total angular momentum at the center of mass \citep{Diaz:2014}. 
On the other hand, it has to be noticed that older results based on the dynamics of M31 dwarf spheroidal companions \citep[e.g.][]{Evans:2000, Gottesman:2002}  
have been pointing to an opposite direction.
Therefore, it needs to be stressed that the choice \mmto$>$\mmw\ is one of the assumptions of this particular Local Group model,
and is not meant to suggest that the ratio of M31 to MW mass ratio is a settled issue.

Since the goal is to derive a posterior distribution function for \mlg, \mmto\ and \mmw, all observational inputs on 
the masses will be neglected. 
The focus will be thus placed on the kinematic properties of the pair: their tangential (\vtan) and radial (\vrad) velocities as well as 
inter-halo separation $r$.
Moreover, even though observations on larger scales (such as the coldness of the local Hubble flow, \citep{Gonzalez:2014}, or the filamentary
nature of our immediate neighbourhood \citep{Libeskind:2015b}) 
have been shown to affect LG properties, environmental considerations are intentionally neglected in our model.
In this way, it will be possible to single out the effects of the LG environment, 
which is by construction reproduced for all of the candidates drawn from the CS sample.
The only external condition which all of the pairs are subjected to is \emph{isolation}, meaning that no other object at least 
as massive as the lightest of the pair must reside within 2.5\hMpc\ from the LG center of mass.

The ranges of allowed values for $r$, \vrad\ and \vtan\ are shown in \Tab{tab:priors}.
A set of six different priors has been used to define separate samples of LG like objects, using two different prescriptions for
$r$ and \vrad, which are very well known quantities, and \vtan.
In the case of the latter, in fact, there is yet no clear consensus on its real value, for which currently two incompatible estimates exist.
These were obtained using different techniques: the one of \citet{Sohn:2012}, who found \vtan$=17\pm17$ (referred to as \vtanI) 
directly measuring the motion stellar population within M31 against background galaxies; and 
the one of \citet{Salomon:2016}, who inferred \vtan=$164\pm62$ (referred to as \vtanII) using precise satellite galaxies data of the
PAndAS survey \citep{McConnachie:2009, Martin:2013a, Martin:2013b}.

Due to the small $1-\sigma$ errors around $r$ and \vrad, 
in order to gather a statistically meaningful sample of LGs 
with properties reasonably close to the
observational ones, we took intervals of $\pm 25\%$ and $\pm 50\%$ of the fiducial values taken from \citet{Marel:2012}.
On the other hand, given the large uncertainties surrounding \vtan, $\pm1\sigma$ priors have been used, to avoid
being unnecessarily restrictive by imposing the same interval prescriptions of the other parameters.
Given these choices, we proceed sampling the posterior distribution functions for 
\mlg, \mmto\ and \mlg\ from both \std\ and CS simulations. % following the procedure of \citet{Busha:2011} and \citet{Gonzalez:2014}.
LG-like pairs are searched for in the full box in the first case, while for each CF2-constrained simulation 
the search volume is limited to a $7$\hMpc\ sphere around the center of the box, 
to ensure that these pairs are living within an environment that
reproduces the main features (such as Virgo, local filament and local void) of the observational one.%, as shown by \citet{Carlesi:2016a}.

\section{Results}
\label{sec:results}
\begin{table*}
\begin{center}
\caption{ 
\label{tab:masses}
Median $\pm$ 75th and 25th percentiles for the masses of the Milky Way, Andromeda and total mass in $10^{12}$\Msun\
units for both constrained and unconstrained simulations. $\mu$ and $\sigma$ represent the  
the best fit values to a lognormal distribution.
Each table shows an estimate given for a different set of priors on velocity (radial and tangential) and relative distance.
The upper three tables implement intervals of $\pm 25\%$ around the fiducial values of \vrad\ and $r$ while those for the lower three
amount to $\pm 50\%$. For each table the number of CS and \std\ pairs per prior choice is also shown.}

\begin{tabular}{rcl}

\begin{tabular}{l|ccc}
\multicolumn{4}{r}{$\qquad$\modo, $N_{CS}=857$, $N_{Rand}=1004$}\\
\hline
$\quad$ &  $M_{200}$ & $\mu$ & $\sigma$ \\
\hline
$M_{M31, CS}$ & $1.77^{+   0.72} _{-   0.91}$ &   12.19 &    0.31\\
$M_{M31, Rand}$ & $1.54^{+   0.55} _{-   0.80}$    &   12.14 &    0.31\\
\hline
$M_{MW, CS}$ & $0.74^{+   0.26} _{-   0.41}$ & 11.80 &    0.30\\ 
$M_{MW, Rand}$ & $0.66^{+   0.28} _{-   0.37}$   &   11.75 &    0.34\\
\hline
$M_{MLG, CS}$  & $2.57^{+   0.87} _{-   1.19}$ &   12.38 &    0.27\\
$M_{MLG, Rand}$ & $ 2.26^{+   0.76} _{-   1.14}$   &   12.30 &    0.27\\
\hline
\end{tabular} & 
\hspace{-0.65cm}

\begin{tabular}{ccc}
\multicolumn{3}{c}{\modtw, $N_{CS}=43$, $N_{Rand}=252$}\\
\hline
 $M_{200}$ & $\mu$ & $\sigma$ \\
\hline
$1.07^{+   0.31} _{-   0.45}$    &   11.93 &    0.21\\
$1.14^{+   0.35} _{-   0.46} $    &   12.00 &    0.25\\
\hline
$0.60^{+   0.21} _{-   0.14}$ &   11.69 &    0.19\\ 
$0.47^{+   0.16} _{-   0.25}$   &   11.62 &    0.26\\
\hline
$1.79^{+   0.51} _{-   0.51}$   &   12.20 &    0.14\\
$1.65^{+   0.44} _{-   0.66}$   &   12.15 &    0.20\\
\hline
\end{tabular} &
\hspace{-0.65cm}

\begin{tabular}{ccc}
\multicolumn{3}{c}{\modth, $N_{CS}=48$, $N_{Rand}=174$}\\
\hline
 $M_{200}$ & $\mu$ & $\sigma$ \\
\hline
$2.48^{+   0.80} _{-   0.56}$ &   12.21 &    0.19\\
$2.93^{+   1.09} _{-   1.57}$  &   12.33 &    0.29\\
\hline
$0.82^{+   0.33} _{-   0.68}$  &   11.76 &    0.35\\ 
$1.06^{+   0.60} _{-   0.83}$  &   11.90 &    0.46\\
\hline
$  3.47^{+   1.16} _{-   0.83}$  &   12.41 &    0.16\\
$  4.13^{+   1.44} _{-   2.11}$  &   12.52 &    0.27\\
\hline
\end{tabular} \\

\begin{tabular}{l|ccc}
\multicolumn{4}{c}{\quad}\\
\multicolumn{4}{c}{\quad}\\
\multicolumn{4}{c}{$\qquad$\modfo, $N_{CS}=1806$, $N_{Rand}=4021$}\\
\hline
$\quad$ & $M_{200}$ & $\mu$ & $\sigma$ \\
\hline
$M_{M31, CS}$ & $1.50^{+   0.72} _{-   0.95}$ &   12.11 &    0.38\\
$M_{M31, Rand}$ & $1.26^{+   0.60} _{-   1.08}$  &   12.08 &    0.40\\
\hline
$M_{MW, CS}$ & $0.58^{+   0.24} _{-   0.44}$  &   11.74 &    0.35\\ 
$M_{MW, Rand}$ & $ 0.51^{+   0.24} _{-   0.45}$ &   11.69 &    0.40\\
\hline
$M_{MLG, CS}$  & $  2.21^{+   0.95} _{-   1.37}$  &   12.29 &    0.34\\
$M_{MLG, Rand}$  & $1.84^{+   0.82} _{-   1.53}$  &   12.24 &    0.37\\
\hline
\end{tabular} & 
\hspace{-0.65cm}

\begin{tabular}{ccc}
\multicolumn{3}{c}{\quad}\\
\multicolumn{3}{c}{\quad}\\
\multicolumn{3}{c}{\modfi, $N_{CS}=160$, $N_{Rand}=949$}\\
\hline
$M_{200}$ & $\mu$ & $\sigma$ \\
\hline
$0.92^{+   0.40} _{-   0.48}$ &   11.93 &    0.31\\
$0.81^{+   0.34} _{-   0.53}$ &   11.88 &    0.34\\
\hline
$0.46^{+   0.19} _{-   0.21}$ &   11.63 &    0.30\\ 
$0.34^{+   0.13} _{-   0.26}$  &   11.53 &    0.31\\
\hline
$1.36^{+   0.54} _{-   0.68}$   &   12.10 &    0.27\\
$1.20^{+   0.49} _{-   0.70}$   &   12.06 &    0.28\\
\hline
\end{tabular} & 
\hspace{-0.65cm}

\begin{tabular}{ccc}
\multicolumn{3}{c}{\quad}\\
\multicolumn{3}{c}{\quad}\\
\multicolumn{3}{c}{\mods, $N_{CS}=125$, $N_{Rand}=551$}\\
\hline
 $M_{200}$ & $\mu$ & $\sigma$ \\
\hline
$ 2.29^{+   0.78} _{-   1.04}$   &   12.28 &    0.28\\
$ 3.00^{+   1.18} _{-   1.65}$   &   12.42 &    0.34\\
\hline
$ 0.79^{+   0.35} _{-   0.65}$   &   11.86 &    0.38\\ 
$ 1.01^{+   0.51} _{-   0.97}$   &   11.96 &    0.42\\
\hline
$ 3.40^{+   1.29} _{-   0.97}$  &   12.46 &    0.26\\
$4.14^{+   1.58} _{-   2.52}$  &   12.57 &    0.33\\
\hline
\end{tabular}

\end{tabular}
\end{center}
\end{table*}

\begin{figure*}
\begin{center}
$\begin{array}{rcl}
\vspace{-0.085cm}
\includegraphics[height=5.4cm, width=6.7cm]{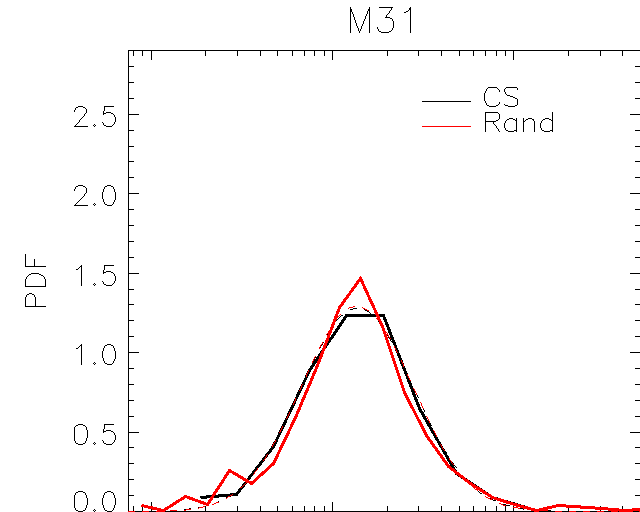} & 
\hspace{-0.7cm}
\includegraphics[height=5.4cm, width=5.5cm]{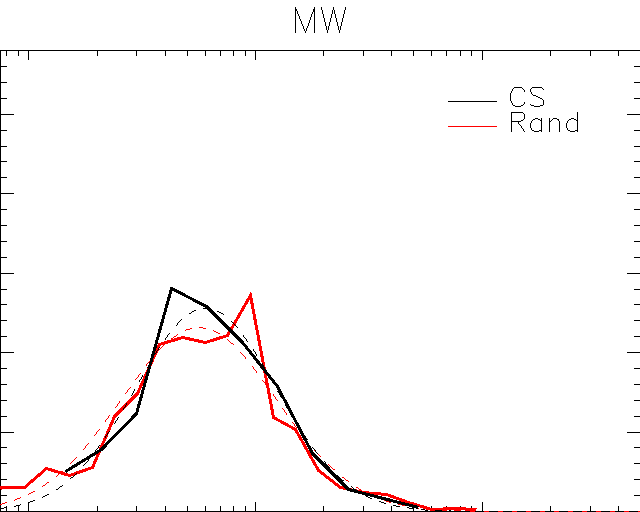} & 
\hspace{-0.7cm}
\includegraphics[height=5.4cm, width=5.9cm]{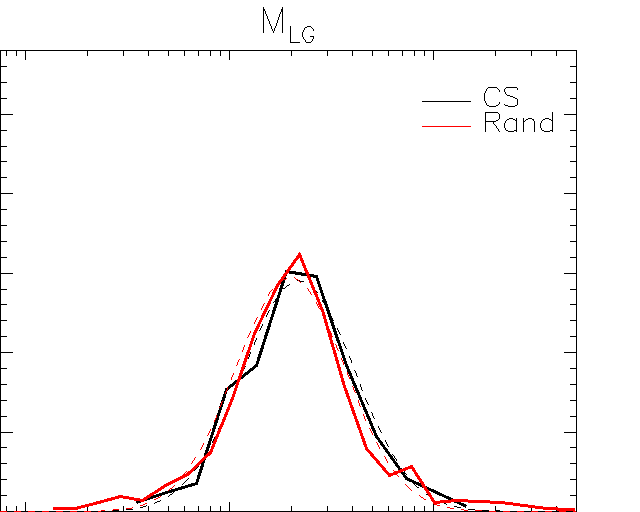} \\

\vspace{-0.075cm}
\includegraphics[height=5.0cm, width=6.7cm]{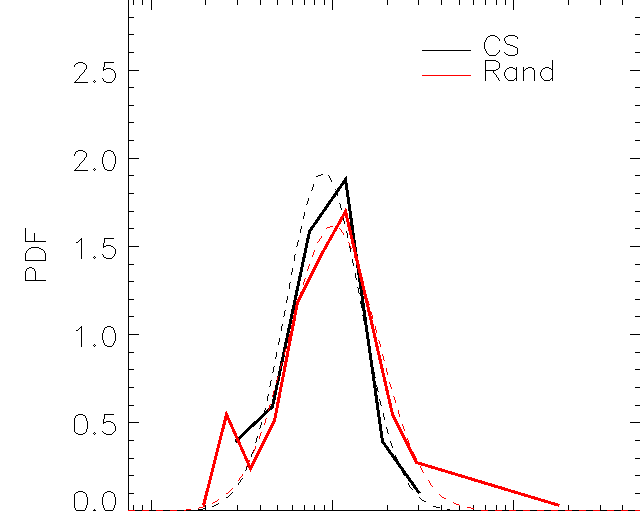} & 
\hspace{-0.7cm}
\includegraphics[height=5.0cm, width=5.5cm]{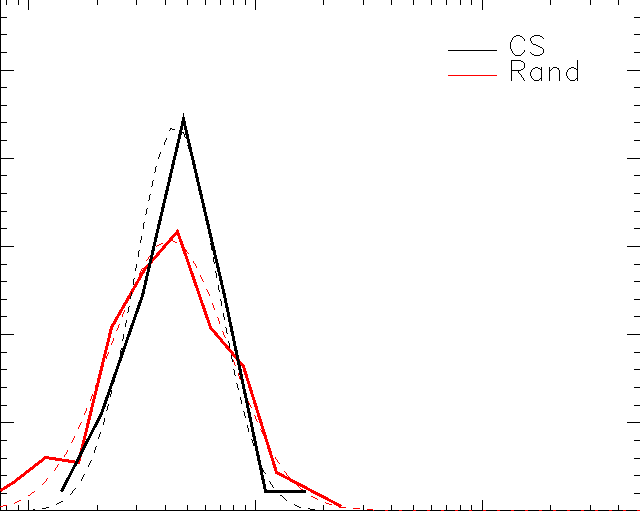} & 
\hspace{-0.7cm}
\includegraphics[height=5.0cm, width=5.9cm]{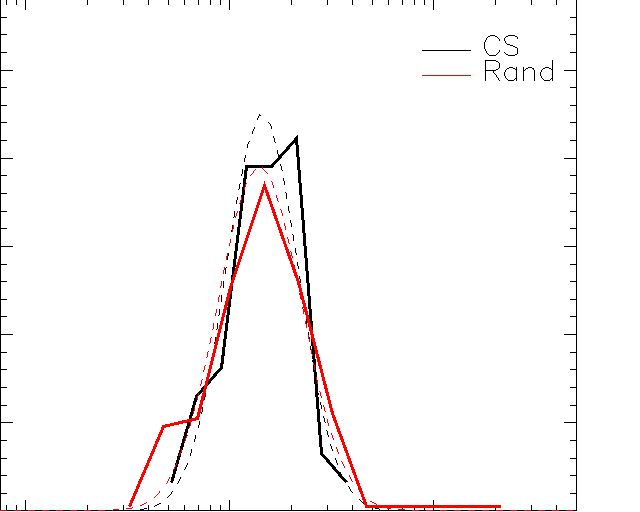} \\

\includegraphics[height=6.0cm, width=6.7cm]{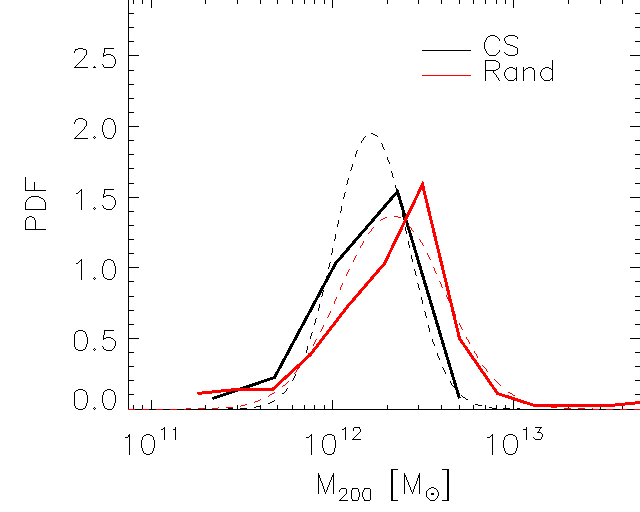} & 
\hspace{-0.7cm}
\includegraphics[height=6.0cm, width=5.5cm]{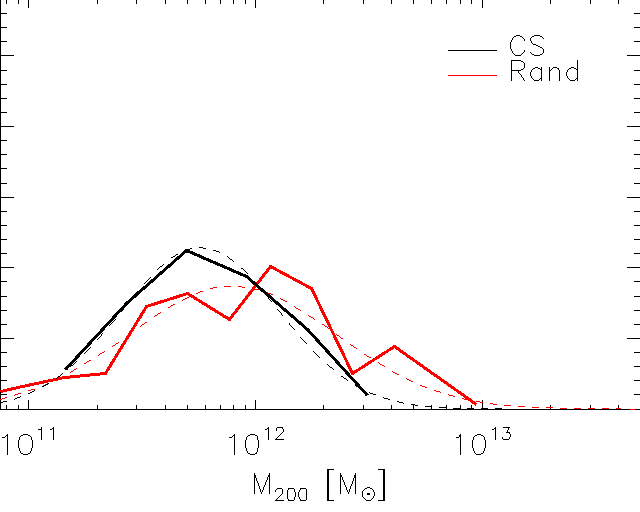} & 
\hspace{-0.7cm}
\includegraphics[height=6.0cm, width=5.9cm]{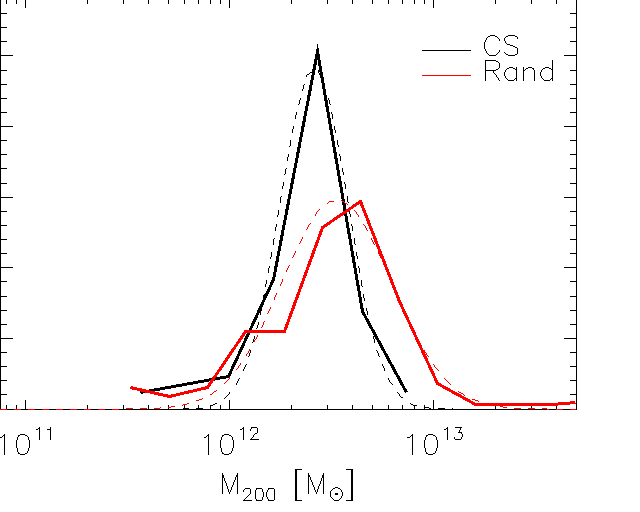} \\
\end{array}$
\caption{Probability distributions for the MW, M31 and LG masses. The figures on the first line show the results 
corresponding to \modo, on the second line to \modtw\ and the third one to \modth.
CS results are characterized by a reduced scatter around the peak likelihood values, as is particularly clear 
in the second and third lines in the case of \mlg\ distributions.
\label{img:distributions}}
\end{center}
\end{figure*}

The six samples corresponding to the parameter combinations of \Tab{tab:priors} are used to derive numerical distributions 
for the masses which can be fitted by a log-normal function (as \citet{Busha:2011} already noted for \mmw):

\begin{equation}\label{eq:gauss}
f(\log_{10}(M_{200}/\Msun)) = \frac{1}{\sqrt{2\pi}\sigma} \exp{-\frac{(\log_{10}(M_{200}/\Msun) - \mu)^2}{2\sigma^2}}
\end{equation}

\noindent
The numerical means and standard deviations for the six combinations of the priors
are shown in \Tab{tab:masses} for the $\log_{10}$ values of the masses in \Msun\ units; while \Fig{img:distributions} shows the 
distributions obtained for the first three sets of parameters.
For each LG candidate we also compute the mass ratio of the larger to the smaller halo of the pair, i.e. \mmto/\mmw\ in our convention, 
and in \Tab{tab:ratios} we provide median, upper and lower quartile values of their distribution for each prior set.
These values can be compared to those shown in \App{app}, which shows a large set of estimates of \mlg, \mmw\ and \mmto\ 
obtained different techniques and under different assumptions.

First of all, it has to be noticed that the $r$ and \vrad\ priors play a minor role in shaping the likelihood function, resulting
in a slight reduction of the peak masses of \mlg, \mmw\ and \mmto\ using $50\%$ intervals at each given \vtan\ prior. 
This finding is in agreement with \citet{Gonzalez:2014} and \citet{Carlesi:2016b}, who found a weak dependence of the posterior distribution function on a 
given set of priors around the fiducial values of \citet{Marel:2012} for $r$ and \vtan.
Therefore, the next subsections will deal only with the results obtained for the first three -- more restrictive -- prior sets,
as our main conclusions will be left largely unaltered by changing them.
We also notice, as a general trend, that CS distributions are characterized by a narrower distribution around the peaks,
leading to smaller error estimates, despite the substantial reduction in the number of objects within the samples.

\subsection{Total mass}
A first look at \mlg\ shows that the choice of the \vtan\ prior plays a main role in shaping the posterior distribution function, as 
can be seen by looking at the values obtained for \vtanI\ and \vtanII, the most extreme cases.
In fact, while the former yields \mlg$^{CS}=   1.79^{+   0.51} _{-   0.52}\times10^{12}$\Msun, the application of the latter results in 
\mlg$^{CS}=   3.47^{+   1.16} _{-   0.83}\times10^{12}$\Msun, a value which is almost a factor of 2 larger and 
nicely overlaps with the number \mlg$=3.6^{+3.0}_{-2.6}\times10^{12}$\Msun\ quoted by \citet{McLeod:2016} feeding the same
kind of \vtanII\ priors to a machine learning algorithm.

This holds also in the case of the \std\ sample, for which similar numerical values have been obtained.
Both results suggest that the total LG should be smaller than those obtained using the TA \citep[e.g.][]{Li:2008, Partridge:2013} 
and least action methods \citep{Phelps:2013}, which generally predict \mlg$>4\times10^{12}$\Msun.
This is most likely due to the approximations behind the TA, as it has been showed that neglecting both \vtan\ \citep{Fattahi:2016a, McLeod:2016} 
and large satellites (such as the Large Magellanic Clouds, see \citet{Penarrubia:2016}) lead to overestimate the value of \mlg.

The above mentioned \mlg\ values are well consistent with the recent findings of \citet{Gonzalez:2014} and \citet{Diaz:2014}, 
who both favour a total mass of $\approx 2.5 \times 10^{12}$\Msun, very close to our \modo\ estimate (which makes no assumptions on \vtan)
that gives \mlg$^{CS}=   2.57^{+   0.87} _{-   1.19}\times10^{12}$. 

So despite the large differences caused by the different tangential velocities, we cannot 
strongly favour or disfavour any among \vtanI\ or \vtanII, as the aforementioned authors place its value between the two.
This means that the \emph{total} \mlg\ alone cannot yet provide strong evidence on the nature of \vtan, and 
knowledge of masses of the individual LG members is required.

We also note that using \vtanI\ priors (or no \vtan\ priors at all), CS mass estimates tend to be larger than \std\ ones.
This is consistent with the findings shown in \citet{Carlesi:2016b}, where it was argued that CS tend to favour higher \vtan s, 
which in turn would lead to higher mass estimates \citep[see][]{Fattahi:2016a}.
Nonetheless, at a first glance, this results seems to be at odds with \citet{Gonzalez:2014}, who pointed out that the total mass of the LG
should be decreasing as a consequence of the environmental constrains on the sample.
However, it has to be emphasized that our choice of the priors on \vtanI\ is much more restrictive ($1\sigma$, \vtan$<34$\kms) 
compared to theirs ($3\sigma$, \vtan$< 68$\kms).
Repeating the analysis using their same prior prescription we found \mlg$^{CS} = 2.12^{+   0.43} _{-   0.40}\times10^{12}$\Msun\ and
\mlg$^{\std}  =2.26^{+   0.40} _{-   0.50}\times10^{12}$\Msun, in good qualitative agreement with their findings.

\subsection{Individual masses}

Our LG model defined M31 (MW) as the heaviest (lightest) halo of the pair, allowing us to inspect the properties of each member of the LG.
Models of the kind \mmw$ > $\mmto\ are neglected in this work, though they cannot be excluded - as noticed earlier.
In what follows, the main findings regarding \mmw\ and \mmto\ are presented along with their consistency with 
existing results found by other authors.\\

\noindent
\textbf{Andromeda} \modo\ and \modth\ results for \mmto\ are in very good agreement with the value of $\log_{10}($\mmto$)=12.3\pm0.1$ derived  
by \citet{Fardal:2013}. Remarkably, although both CS and \std\ values overlap with it within the scatter, 
best fit peak values for the former group of simulations are remarkably much closer.
Indeed, if M31 turns out to be this massive, it would directly favour the \vtanII\ value over \vtanI, in agreement 
with \citet{Fattahi:2016a} who pointed out that larger masses are in general associated with higher tangential velocities in
cosmological simulations.
On the other hand, it turns out the likelihood functions of \mmto\ derived with small tangential velocity LG-model 
(\modtw), show a well defined peak around a smaller mass, \mmto$^{CS}=   1.07^{+   0.27} _{-   0.50}\times10^{12}$\Msun.
Such a light M31 would still be compatible with some of the older estimates \citep[see e.g.][]{Evans:2000, Gottesman:2002, Ibata:2004}, 
for \vrad\ and $r$, 
but is of course in conflict with the aforementioned -- more recent -- results, reminding us of the importance of a precise
measurement of \vtan\ in oder to obtain a complete and consistent picture of the LG. \\

\noindent
\textbf{Milky Way} As a consequence of our initial assumption \mmw$<$\mmto, the MW is characterized by a small mass.
This is true in particular in the CS case, for which \vtanI\ values yield \mmw$_{CS}=   0.60^{+   0.21} _{-   0.14}\times10^{12}$\Msun\ and 
\mmw$_{CS}=   0.82^{+   0.33} _{-   0.68}\times10^{12}$\Msun\ for \vtanII. These numbers are both consistent with a large number of recent results
\citep{Bovy:2012, Deason:2012, Rashkov:2013, Gibbons:2014, Eadie:2016} that indicate a MW mass in the range $(0.5 - 1.0) \times 10^{12}$\Msun.
On the other hand, they appear to be on the low end of the 
$1.26\pm0.24\times10^{12}$\Msun\ and 
$1.3 \pm 0.3 \times 10^{12}$\Msun\ values computed by \citet{McMillian:2011, McMillian:2016} modelling the mass 
distribution of the Galaxy, as well as the $1.2-1.7\times10^{12}$\Msun\ 
constraints of \citet{Fragione:2016} obtained using ultra high velocity stars.
Moreover, such a light mass would in turn disfavour a bound orbit for Leo I.
This remark is consistent with the analysis made by \citet{Watkins:2010} and \citet{Boylan-Kolchin:2013}, who showed that a bound
Leo I alone would shift the \mmw\ estimate upwards by $\approx 30\%$.
Therefore, removing it from the list of MW satellites, significantly smaller \mmw\ values can be obtained, leading to a good agreement with our results.
On the other hand, an unbound Leo I at odds with what has been suggested by other authors, most notably \citet{Boylan-Kolchin:2013}, arguing in 
favour of a bound orbit based on $N$-body simulations.
However, in order to consistently address the issue at the root of this discrepancy from the standpoint of CSs, 
it is necessary to use higher resolution simulations, which allow to deal with substructure with a sufficient level of accuracy;
a subject which is currently being pursued. \\

\begin{table}
\begin{center}
\caption{Median ratios \mmto/\mmw\ for the different LG models, within CS and \std\ LG samples. Intervals are given by the 25-th and 75-th percentiles
of the distributions. Lower \vtan\ values are correlated with smaller mass ratios in both CS and \std. 
We also note that, for any given set of priors, CS samples predict smaller values of \mmto\ to \mmw\ with a reduced scatter around the median value. }
\label{tab:ratios}\begin{tabular}{ccc}
\hline
Prior & CS  & \std \\ 
\hline
\modo & $   1.90^{+   1.25} _{  -0.58}$ & $   2.10^{+   1.67} _{  -0.70}$\\
\quad & \quad & \quad \\
\modtw & $   1.64^{+   0.61} _{  -0.34}$ & $   2.03^{+   1.16} _{  -0.79}$\\
\quad & \quad & \quad \\
\modth & $   2.01^{+   1.88} _{  -0.66}$ & $   2.13^{+   2.87} _{  -0.89}$\\
\quad & \quad & \quad \\
\modfo & $   1.90^{+   1.43} _{  -0.57}$ & $   2.09^{+   1.79} _{  -0.69}$\\
\quad & \quad & \quad \\
\modfi & $   1.73^{+   0.83} _{  -0.40}$ & $   1.96^{+   1.41} _{  -0.58}$\\
\quad & \quad & \quad \\
\mods & $   2.13^{+   1.66} _{  -0.76}$ & $   2.28^{+   2.19} _{  -1.01}$\\
\hline
\end{tabular}
\end{center}
\end{table}

\noindent
\textbf{Mass ratios} Given the above results for the individual halo masses, it is possible to draw a few final conclusions regarding the
\emph{mass ratios} of the pair, which were shown in \Tab{tab:ratios}.
The numerical values of these ratios show a good agreement with results obtained by
\citet{Diaz:2014}, who derived a ratio of $2.3^{+2.1}_{-1.1}$ by enforcing the zero angular momentum condition 
with respect to the center of mass of the LG.
Lower ratios, such as the one of $\approx 5/4$ derived by \citet{Karachentsev:2009},
seem to be slightly in tension with the present results, which favour ratios $> 1.3$ for both CS and \std\ simulations at the $75\%$ level.
Thanks to the use of different LG models and different halo samples it is possible to disentangle 
the influence that either of the two play in determining those ratios.
%\noindent
We first remark that both CS and \std\ LGs show the same qualitative behaviour under a change of \vtan\ priors:
namely, the gap between MW and M31 masses is enhanced when switching from \vtanI\ to \vtanII\ constraints.
In particular, this change is larger (shifting up to $25\%$, from 1.64 to 2.01) in the case of CS, 
as a consequence of the minor change of \mmw\ masses (from $0.6\times10^{12}$\Msun to $0.8\times10^{12}$\Msun) 
with respect to \mmto, which increases from $\approx1.0\times10^{12}$\Msun\ to $\approx2.0\times10^{12}$\Msun\ with increasing \vtan.
Ratios in the \std\ sample, though affected in the same manner, appear less pronounced and are characterized by larger variability intervals; moreover,
the median values of these ratios are always above the CS ones, overlapping at the confidence level of the 25-th and 75-th percentiles.
It is thus possible to conclude that \vtan\ values not only affect individual masses, but mass ratios as well, by affecting
them unevenly. 

%%%%%%%%%%%%%%%%%%%%%%%%%%%%%%%%%%%%%%%%%%%%%%%%%%%%%%%%%%%%%%%%%%%%%%%%%%%%%%%%%%

\section{Conclusions}\label{sec:conclusions}
We have derived the posterior probability distribution functions of the mass of the Local Group and its main constituents, the Milky Way and the M31 galaxies. 
These posterior distributions have been derived by means of constrained and random DM-only cosmological simulations. 
The prior assumptions that condition the above probability distributions are: 
a. The standard \LCDM\ cosmological model; 
b. A Local Group model, which summarizes the observational knowledge of the LG and thereby defines what a LG-like object is. 
The numerical simulations are the means by which the posterior distributions are constructed - 
the \LCDM\ model is used to set up the simulations and the LG model is used to select LG-like objects. 
The Cosmicflows-2 database of peculiar velocity is setting another prior on the 
constrained simulations, which are designed to reproduce the imprint of the local 
universe on the CF2 data.   
The choice of the LG model and the CF2 constraint allows us to study the way 
the structure of the LG and the nearby universe affects our knowledge of masses 
of the MW and M31.
A particular emphasis has been placed also on the effects of using a low-\vtan\ estimate 
(the value found by \citet{Sohn:2012}, referred to as \vtanI) and a high one 
(\vtanII, derived by \citet{Salomon:2016}). 

We applied these models to objects drawn from two different types of simulations: a series of constrained simulations
(CS), from which it was possible to gather halo pairs living in a large scale environment akin to the observational one, and a standard, random \LCDM\ (\std).
We derived posterior distribution functions for the mass of the Local Group (LG) and its most prominent members, MW and M31,
using the cosmological model \LCDM\ and different possible definitions for the LG iself.
This setup was designed to address several questions:

\begin{itemize}
\item what are the implications of the different \vtan\ values for \mlg, \mmto\ and \mmw? 
\item what is the role of the environment in shaping the posterior distribution functions?
\item how do different priors on well measured quantities (\vrad\ and $r$) affect the predicted mass distribution?
\end{itemize}

After establishing that the particular choices for \vrad\ and $r$ priors have only a minor effect
on the results (in agreement with the findings of \citet{Gonzalez:2014, Carlesi:2016b}), it has been possible to determine the following: \\

\noindent
(a) \vtan\ priors play a dominant role in shifting the peak of the likelihood for all the masses. In fact, estimates based on \vtanI\ and \vtanII\ 
differ up to more than a factor of 2.\\

\noindent
(b) CS estimates for masses and ratios are characterized by a $\approx 25\%$ less in the scatter around the median values compared to \std\ results. \\

\noindent
(c) \mmto\ to \mmw\ ratios are slightly but systematically lower in CSs than in their \std\ counterparts. Moreover, they are more affected by the change
in \vtan\ prior choice, with a $\approx 25\%$ difference between \vtanI\ and \vtanII. \\

\noindent
(d) MW mass values derived from CS samples consistently point towards \mmw s $\approx (0.5 - 1.0) \times 10^{12}$\Msun, in good agreement with the recent 
findings of \citet{Deason:2012, Bovy:2012, Rashkov:2013, Gibbons:2014}, and lying on the lower end of the intervals provided by \citet{McMillian:2011, 
Fragione:2016, McMillian:2016}. This is largely independent of the choice of \vtan.\\

\noindent
(e) Peak likelihood masses for M31 in CS simulation (\modo\ and \modth) are extremely close to the $\log_{10}$\mmto$ = 12.3 \pm 0.1$ value \citet{Fardal:2013}, 
indicating that a higher \vtan\ value is in better agreement with those data. 
\std\ results are also compatible within the errors with those estimated, though their peak likelihoods are not as close.
\\

\noindent
We also found that our \mlg\ estimates tend to be smaller than TA ones, 
in agreement with \citet{Gonzalez:2014, Fattahi:2016a}).
This inconsistency cannot be explained by definitions of the mass alone (e.g. collapsed mass vs. $M_{vir}$) and is
most likely due to the crude approximations of the method \citep{Penarrubia:2016}.
On the other hand, the results we obtained for the masses significantly overlap with previous estimates obtained by other means, 
while at the same time providing more insight and singling out the effects of the environment and the implications of each different \vtan\ measure. 
Moreover, the scatter around the peaks for the mass likelihood functions turns out to be smaller as a consequence of the 
constrained variance, as opposed to the unconstrained results which display consistently higher $\sigma$s.

\noindent
Future investigation along these lines will be based on a set of higher resolution Local Group Factory simulations, which will
allow to study mass accretion histories and satellite populations.
In particular, we plan to analyze the properties of substructure around MW and M31 analogues
to clarify the effect of the farthest massive satellites on LG mass estimation and address the source of
disagreement with the MW mass estimates of \citet{Busha:2011} and \citet{Boylan-Kolchin:2013}, 
which were inferred by the properties of the Large Magellanic Cloud and Leo I, respectively.

To sum up, we have shown that our Bayesian analysis of the masses of the local group and its main members
provides consistent results given a simple prior model (our LG model and \LCDM) and CF2 data.
Moreover, we have been able to derive useful formulas that can be generally used for analytical modelling of the LG masses.

\section*{Acknowledgements}

EC would like to thank the Lady Davis Fellowship Fund for financial support. 
YH has been partially supported by the Israel Science Foundation (1013/12).
JS acknowledges support from the Alexander von Humboldt foundation.
%SG and YH acknowledge support from DFG under the grant GO563/21-1.
%GY thanks MINECO (Spain) for financial support under  project grants AYA2012-31101 and AYA2015-63810-P.
%We thank the anonymous referee for his useful remarks.
%We thank the Red Espa\~nola de Supercomputaci\' on   for  granting us 
%computing time  in the Marenostrum Supercomputer at the BSC-CNS where 
%the Local Group Factory simulations have been performed, 
%as well as PRACE for granting computing time in the CURIE supercomputer  where the CurieHZ simulations was ran.

%%%%%%%%%%%%%%%%%%%%%%%%%%%%%%%%%%%%%%%%%%%%%%%%%%%

\bibliographystyle{mn2e}
\bibliography{biblio}

\bsp

%%%%%%%%%%%%%%%%%%%%%%%%%%%%%%%%%%%%%%%%%%%%%%%%%%%

\appendix
\section{Local group masses}
\label{app}

In this appendix we provide a summary of some different estimates of \mmw\ (shown in \Tab{tab:appmw}), \mmto\ (\Tab{tab:appmmto}) and \mlg\ (\Tab{tab:appmlg}).
These tables contain mostly those values which have been quoted throughout this work and is provided as 
a quick reference for the readed to the vast amount of results obtained using a large number of techniques. 
Due to the huge number of works that can be found in the literature on the subject, these tables are not meant to be exhaustive accounts 
of all the results, but rather present a broad overview of the variety of the methods used and values that have been obtained.
For some authors more than one value is quoted when different assumptions were used within the same publication.\\

\begin{table*}
\begin{center}
\caption{Milky Way mass estimates in $10^{12}$\Msun\ units.}
\label{tab:appmw}
\begin{tabular}{cccc}
\multicolumn{4}{c}{\mmw}\\
\hline
Source & Type of mass & Value & Method \\
\hline
\citet{Kochanek:1996} & $M(100 \kpc)$ & $0.54\pm0.13$ & Satellites \& stars escape velocity \\
\citet{Kochanek:1996} & $M(100 \kpc)$ & $0.33 - 0.61$ & Satellites \& stars escape velocity incl. Leo I\\
\citet{Wilkinson:1999} & \mvir & $1.9^{+3.6}_{-1.7}$ & Satellites \& globular clusters \\
\citet{Klypin:2002} & \mvir & $1.0-2.0$ & Cuspy halo, adiabatic contraction \\
\citet{Sakamoto:2002} & \mvir & $2.5^{+0.5}_{-1.0}$ & Satellites (incl. Leo I) \\
\citet{Sakamoto:2002} & \mvir & $1.8^{+0.0}_{-0.7}$ & Satellites (excl. Leo I) \\
\citet{Battaglia:2005} & \mvir & $0.8^{+1.2}_{-0.2}$ & Radial velocity dispersion \& NFW profile \\	
\citet{Battaglia:2005} & \mvir & $1.2^{+1.8}_{-0.5}$ & Radial velocity dispersion \& Truncated flat profile \\	
\citet{Dehnen:2006} & \mvir & $\approx1.5$ & Radial velocity dispersion \\
\citet{Smith:2007} & \mvir & $1.42^{+1.14}_{-0.54}$ & Local galactic escape speed \\
\citet{Xue:2008} & \mvir & $1.0^{+0.3}_{-0.2}$ & Line-of-sight velocity distribution \\	
\citet{Li:2008} & \mth & $2.43^{+0.60}_{-0.73}$ & TA calibrated on simulations \\
\citet{Gnedin:2010} & $M(<80 \kpc)$ & $0.69^{+0.30}_{-0.12}$ & Halo stars radial velocities \\	
\citet{McMillian:2011} & \mvir & $1.26\pm0.24$ & Satellites \& radial DM distribution models \\ 
\citet{Busha:2011} & \mvir & $1.2^{+0.7}_{-0.4}$  & Large Magellanic clouds motion\\
\citet{Bovy:2012} & \mvir & $\approx 0.8$ & Rotation curve \\
\citet{Deason:2012a} & $M(<50\kpc)$ & $\approx 0.4$ & Blue horizontal branch star kinematics \\
\citet{Deason:2012} & $M(<150\kpc)$ & $(0.5-1.0)$ & Radial velocities of stellar halo \\
\citet{Boylan-Kolchin:2013} & \mvir & $1.6^{+0.8}_{-0.6}$ & Leo I motion\\
\citet{Rashkov:2013} & \mvir & $\approx 0.8$ & SDSS halo stars \& numerical simulations \\
\citet{Gonzalez:2013} & \mth & $1.14^{+1.19}_{-0.41}$  & Large Magellanic clouds properties\\
\citet{Gibbons:2014} & $M(<200\kpc)$ & $0.56\pm0.12$ & Sagittarius stream \\
\citet{Diaz:2014} & \mvir & $0.8\pm0.5$ & Momentum balance at the LG center of mass \\
\citet{Cautun:2014} & \mth & $0.25 - 1.4$ & Maximum circular velocity of satellites\\
\citet{McMillian:2016} & \mvir & $1.30\pm0.30$ & Satellites \& radial DM distribution models \& gas discs \\ 
\citet{Fragione:2016} & \mvir & $1.2 - 1.7$ & Hypervelocity stars \\
\citet{Eadie:2016} & $M(<179 \kpc)$ & $0.56 - 0.67$ & Bayesian analysis of globular clusters \\
\citet{Penarrubia:2016} & \mvir & $1.04^{+0.26}_{-0.23}$ & TA \& Large Magellanic Clouds corrections\\
\citet{Penarrubia:2016b} & \mvir & $\approx0.85$ & Spherical collapse model \& local Hubble flow \\
\citet{Zaritsky:2016} & \mvir & $> 0.77$ & Universal baryon fraction \\
Present paper & \mth & $0.60^{+0.21}_{-0.14}$ & CS \& \vtanI \\
Present paper & \mth & $0.82^{+0.33}_{-0.68}$ & CS \& \vtanII \\
\hline
\end{tabular}
\end{center}
\end{table*}

\begin{table*}
\begin{center}
\caption{Andromeda mass estimates in $10^{12}$\Msun\ units.}
\label{tab:appmmto}
\begin{tabular}{cccc}
\multicolumn{4}{c}{\mmto}\\
\hline
Source & Type of mass & Value & Method \\
\hline
\citet{Evans:2000} & \mvir & $1.23^{+1.8}_{-0.6}$ & Satellites \\
\citet{Klypin:2002} & \mvir & $1-2$ & Cuspy halo model \& adiabatic contraction \\
\citet{Gottesman:2002} & $M(<350\kpc)$ & $ < 0.6 $ & Dwarf satellites kinematics \\
\citet{Ibata:2004} & $M(<125 \kpc)$ & $0.75^{+0.25}_{-0.13}$ & Giant stream kinematics \\
\citet{Watkins:2010} & $M(<300 \kpc)$ & $1.40\pm{0.43}$ & Satellites \\
\citet{Tollerud:2011} & \mvir & $1.2^{+0.9}_{-0.7}$ & Satellites \\
\citet{Fardal:2013} & \mth & $1.99^{+0.51}_{-0.41}$ & Giant stream kinematics \\
\citet{Diaz:2014} & \mvir & $1.7\pm0.3$ & Momentum balance at the LG center of mass \\
\citet{Penarrubia:2016} & \mvir & $1.33^{+0.39}_{-0.33}$ & TA \& Large Magellanic Clouds corrections\\
Present paper & \mth & $1.07^{+0.31}_{-0.45}$ & CS \& \vtanI \\
Present paper & \mth & $2.48^{+0.80}_{-0.56}$ & CS \& \vtanII \\
\hline
\end{tabular}
\end{center}
\end{table*}

\begin{table*}
\begin{center}
\caption{Local Group mass estimates in $10^{12}$\Msun\ units.}
\label{tab:appmlg}
\begin{tabular}{cccc}
\multicolumn{4}{c}{\mlg}\\
\hline
Source & Type of mass & Value & Method \\
\hline
\citet{Courteau:1999} & $M(< 1.18 \Mpc)$ & $2.3\pm0.6$ & Velocity dispersion \\
\citet{Li:2008} & \mth & $5.27^{+0.93}_{-1.91}$ & TA calibrated on simulations \\
\citet{Marel:2008} & \mvir & $5.58^{+0.85}_{-0.72}$ & TA \\
\citet{Karachentsev:2009} & $M(<0.96 \Mpc)$ & $1.9\pm0.2$ & Local Hubble flow measurement \\
\citet{Marel:2012} & \mvir & $4.93\pm1.63$ & TA \\
\citet{Marel:2012} & \mvir & $3.17\pm0.57$ & Bayesian analysis \\
\citet{Partridge:2013} & \mvir & $4.73\pm1.03$ & TA \& $\Lambda$ corrections\\
\citet{Diaz:2014} & \mvir & $2.4\pm0.8$ & Momentum balance at the LG center of mass \\
\citet{Gonzalez:2014} & \mth & $2.40^{+0.55}_{-0.36}$ & Likelihood estimate \\
\citet{Gonzalez:2014} & $M(<1 \Mpc)$ & $4.17^{+1.45}_{-0.93}$ & Likelihood estimate \\
\citet{Penarrubia:2016} & \mvir & $2.64^{+0.42}_{-0.38}$ & TA \& Large Magellanic Clouds corrections\\
\citet{McLeod:2016} & \mvir & $4.9^{+0.8+1.3}_{-0.8-1.4}$ & Machine learning \& \vtanI \\
\citet{McLeod:2016} & \mvir & $3.6^{+1.3+1.7}_{-1.1-1.5}$ & Machine learning \& \vtanII \\
Present paper & \mth & $1.79^{+0.51}_{-0.51}$ & CS \& \vtanI \\
Present paper & \mth & $3.47^{+1.16}_{-0.83}$ & CS \& \vtanII \\
\hline
\end{tabular}
\end{center}
\end{table*}

\label{lastpage}

\end{document}